# Canted ferrimagnetism and giant coercivity in the non-stoichiometric double perovskite La$_2$Ni$_{1.19}$Os$_{0.81}$O$_6$

Hai L. Feng[1], Manfred Reehuis[2], Peter Adler[1], Zhiwei Hu[1], Michael Nicklas[1], Andreas Hoser[2], Shih-Chang Weng[3], Claudia Felser[1], Martin Jansen[1]

[1]*Max Planck Institute for Chemical Physics of Solids, Dresden, D-01187, Germany*
[2]*Helmholtz-Zentrum Berlin für Materialien und Energie, Berlin, D-14109, Germany*
[3]*National Synchrotron Radiation Research Center (NSRRC), Hsinchu, 30076, Taiwan*

Abstract:
The non-stoichiometric double perovskite oxide La$_2$Ni$_{1.19}$Os$_{0.81}$O$_6$ was synthesized by solid state reaction and its crystal and magnetic structures were investigated by powder *x*-ray and neutron diffraction. La$_2$Ni$_{1.19}$Os$_{0.81}$O$_6$ crystallizes in the monoclinic double perovskite structure (general formula *A*$_2$*BB*'O$_6$) with space group *P*2$_1$/*n*, where the *B* site is fully occupied by Ni and the *B*' site by 19 % Ni and 81 % Os atoms. Using *x*-ray absorption spectroscopy an Os$^{4.5+}$ oxidation state was established, suggesting presence of about 50 % paramagnetic Os$^{5+}$ (5$d^3$, *S* = 3/2) and 50 % non-magnetic Os$^{4+}$ (5$d^4$, *J*$_{eff}$ = 0) ions at the *B*' sites. Magnetization and neutron diffraction measurements on La$_2$Ni$_{1.19}$Os$_{0.81}$O$_6$ provide evidence for a ferrimagnetic transition at 125 K. The analysis of the neutron data suggests a canted ferrimagnetic spin structure with collinear Ni$^{2+}$ spin chains extending along the *c* axis but a non-collinear spin alignment within the *ab* plane. The magnetization curve of La$_2$Ni$_{1.19}$Os$_{0.81}$O$_6$ features a hysteresis with a very high coercive field, *H*$_C$ = 41 kOe, at *T* = 5 K, which is explained in terms of large magnetocrystalline anisotropy due to the presence of Os ions together with atomic disorder. Our results are encouraging to search for rare earth free hard magnets in the class of double perovskite oxides.

## I. INTRODUCTION

Double perovskite oxides $A_2BB'O_6$ containing $3d$-$5d$($4d$) transition metal ions at the $B$ and $B'$ sites are attracting great attention, due to their interesting physical properties. High-temperature ferrimagnetic half-metals, ferro- or ferrimagnetic insulators, as well as materials with large magnetoresistance and exchange bias, are found in this class of compounds [1–9]. They thus show prospects for spintronic applications. Most of these compounds contain $5d$ elements with $5d^2$ and $5d^3$ electronic configurations. Here, the interesting properties arise from competing interactions within and between the $3d$ and $5d$ sublattices. Studies on $5d^4$ systems are rare, probably because of the expected nonmagnetic ground state for $5d^4$ ions. In the presence of strong spin-orbit coupling (SOC), the orbital angular momentum associated with the three $t_{2g}$ orbitals in an octahedral crystal field ($l$ = 1) entangles with the spin moments of the electrons which results in an upper $j_{eff}$ = 1/2 doublet and a lower $j_{eff}$ = 3/2 quadruplet. The formation of a SOC-assisted Mott insulating state in $Sr_2IrO_4$ ($Ir^{4+}$, $t_{2g}^5$) is understood within this picture [10]. According to this scenario, the ground state of a $5d^4$ system in an octahedral crystal field is a trivial singlet with four electrons filling the lower quadruplet, resulting in a nonmagnetic state with total angular momentum $J_{eff}$ = 0. Many studies focus on $Ir^{5+}$ compounds, such as $A_2BIrO_6$ ($A$ = Sr, Ba; $B$ = Sc, Y) [11–17]. In contrast to the expectations for $J_{eff}$ = 0, long-range magnetic order at 1.3 K was reported for $Sr_2YIrO_6$ [11,12] and considered as evidence for intermediate-strength rather than strong spin-orbit coupling which is usually anticipated for $5d$ ions. These results, however, have been challenged and the observed magnetism in such materials may arise from defects and/or nonstoichiometry [13,18].

Beside $Ir^{5+}$, also $Os^{4+}$ ions adopt the $5d^4$ electronic configuration. The singlet ground-state magnetism was discussed in $AOsO_3$ ($A$ = Ca, Sr, Ba) [19] and $R_2Os_2O_7$ ($R$ = Y, Ho) [20]. To the best of our knowledge, the magnetic properties of double perovskites containing $Os^{4+}$ have not been reported yet. In this work, targeting at an $Os^{4+}$ double perovskite $La_2NiOsO_6$, we synthesized a nonstoichiometric phase $La_2Ni_{1.19}Os_{0.81}O_6$ with $Ni^{2+}$ ions at the $B$ site and a mixture of $Ni^{2+}$, $Os^{4+}$, and $Os^{5+}$ ions at the $B'$ site of the double perovskite structure. It was verified by x-ray absorption spectroscopy that the valence state of Os is about 4.5+. Magnetization and neutron diffraction studies revealed that $La_2Ni_{1.19}Os_{0.81}O_6$ shows a ferrimagnetic transition at 125 K and features an extraordinary broad hysteresis with a giant coercivity of 41 kOe at 5 K. These results are encouraging for the search of hard magnetic materials in the class of $3d$/$5d$ double perovskites.

## II. EXPERIMENTAL

A polycrystalline sample of $La_2Ni_{1.19}Os_{0.81}O_6$ was synthesized by solid-state reaction from $La_2O_3$, NiO, and Os. $La_2O_3$, NiO, and Os in a molar ratio 1 : 1 : 1 were well ground together and pressed into a pellet. This was loaded into a corundum crucible which was placed into a silica tube along with a second corundum crucible containing $MnO_2$ (Alfa 99.9%). The silica tube was then sealed under dynamic vacuum using a $H_2/O_2$ torch, and heated at 1250 °C for 48 hours in a tube furnace. $MnO_2$ decomposes into ½ $Mn_2O_3$ + ¼ $O_2$ at 550 °C and acts as an oxygen source for the reaction. The molar ratio of Os and $MnO_2$ is 1 : 4. Small pieces of $La_2Ni_{1.19}Os_{0.81}O_6$ were cut from the synthesized pellet and finely ground to a fine powder, which was characterized by powder *x*-ray diffraction using a HuberG670 camera [Guinier technique, $\lambda$ = 1.54056 Å (Cu-K$\alpha_1$)]. A powder pattern was collected in the 2θ range between 10.9 and 100.2°. A scanning electron microscope (SEM, Philips XL30) with an attached energy dispersive *x*-ray spectrometer (EDX) was used for elemental analysis.

The Os-$L_3$ XAS spectra of $La_2Ni_{1.19}Os_{0.81}O_6$, an $Os^{5+}$ $Sr_2FeOsO_6$ reference, and an $Os^{4+}$ reference $La_2MgOsO_6$ were measured in transmission geometry at the beamline BL07A at the National Synchrotron Radiation Research Center in Taiwan.

Neutron powder diffraction experiments of $La_2Ni_{1.19}Os_{0.81}O_6$ were carried out on the instruments E2, E6, and E9 at the BER II reactor of the Helmholtz-Zentrum Berlin. The instrument E9 uses a Ge monochromator selecting the neutron wavelength $\lambda$ = 1.309 Å, while the instruments E2 and E6 use a pyrolytic graphite (PG) monochromator selecting the neutron wavelength $\lambda$ = 2.38 and 2.417 Å, respectively. In order to investigate in detail the crystal structure of $La_2Ni_{1.19}Os_{0.81}O_6$ at 3.2 and 160 K, well below and above the magnetic ordering temperature, neutron powder patterns were recorded on the instrument E9 between the diffraction angles 7.5 and 141.8°. For a detailed analysis of the magnetic structure and its temperature dependence, we have collected powder patterns between 1.7 and 134 K on the instrument E6 between the diffraction angles 5 and 136.4°. In addition, powder patterns at 1.7 K and 150 K with higher counting rate and better instrumental resolution in the 2θ range between 11.8 and 87.3° were obtained on the instrument E2.

Rietveld refinements of the powder diffraction data were carried out with the program *FullProf* [21]. In the case of the data analysis of *x*-ray diffraction data, we used the atomic scattering form factors provided by the program. For the refinements of the neutron powder data the nuclear scattering lengths $b$(O) = 5.803 fm, $b$(La) = 8.24 fm, $b$(Ni) = 10.3 fm, and $b$(Os) = 10.7 fm were used [22]. The magnetic form factors of the Fe and Os atoms were taken from Refs. [23,24].

Using a polycrystalline piece, the electrical resistivity ($\rho$) was measured with direct current (0.1 mA) in a four-point in-line arrangement (PPMS, Quantum Design). Electrical contacts were made with Au wires and silver paste. The temperature dependence of the magnetic susceptibility was measured in a SQUID magnetometer (MPMS-5T, Quantum Design). The measurements were conducted in warming after zero-field cooling (ZFC) and during field-cooling (FC) in the temperature range 2 – 300 K under applied magnetic fields of 10 kOe. The high-temperature magnetic susceptibility (300 – 565 K) was measured using the same SQUID magnetometer with oven option. Isothermal magnetization curves were initially recorded for fields up to ±50 kOe at temperatures of 5 and 150 K using the same SQUID magnetometer. As the magnetization curves turned out to be unsaturated at 5 K and showed a broad hysteresis, $M(H)$ curves were further measured for fields up to ±140 kOe at 5 K using a Physical Property Measurement System (PPMS, Quantum Design).

## III. RESULTS

**A Crystal structure**

The room-temperature crystal structure of La$_2$Ni$_{1.19}$Os$_{0.81}$O$_6$ was investigated by $x$-ray powder diffraction as shown in Figure 1 (top). La$_2$Ni$_{1.19}$Os$_{0.81}$O$_6$ was successfully refined in the monoclinic space group $P2_1/n$ (No. 14, standard setting $P2_1/c$). In this space group, the La and the three O atoms (O1, O2, and O3) occupy the Wyckoff position 4$e$($x,y,z$), while the Ni and Os atoms are at the positions 2$c$(½,0,½) and 2$d$(½,0,0), respectively. During the refinement, we used the constraint $occ$(Ni) + $occ$(Os) = 1 at the 2$c$ and 2$d$ sites. The refinements indicated that the sample is nonstoichiometric compared to the ideal composition La$_2$NiOsO$_6$ and the following occupancies were found: $occ$(Ni1)/$occ$(Os1) = 0.187(3)/0.813(3) at 2$c$, and $occ$(Ni2)/$occ$(Os2) = 1.001(2)/ −0.001(2) at 2$d$. This suggests that the 2$d$ site is fully occupied by Ni, and therefore in the final refinement, the Ni occupancy at the 2$d$ site was fixed to be 1. This gives the composition La$_2$Ni$_{1.187(3)}$Os$_{0.813(3)}$O$_6$. The refinements resulted in a residual $R_F$ = 0.0180 (defined as $R_F$ = $\sum||F_{obs}| − |F_{calc}||/\sum|F_{obs}|$). The sample contains a smaller amount of La$_3$OsO$_7$ (5.6 %) [25,26]. For this compound, only the overall scale factor and the lattice parameters were simultaneously allowed to vary during the Rietveld refinements, whereas the structural parameters, taken from Ref. 25, were fixed. The loss of Os is presumably due to the formation of volatile $OsO_4$ during the synthesis. The EDX analysis revealed that the sample contains much more Ni than Os, Ni/Os = 1.17/0.83, which is close to the results from the Rietveld refinement.

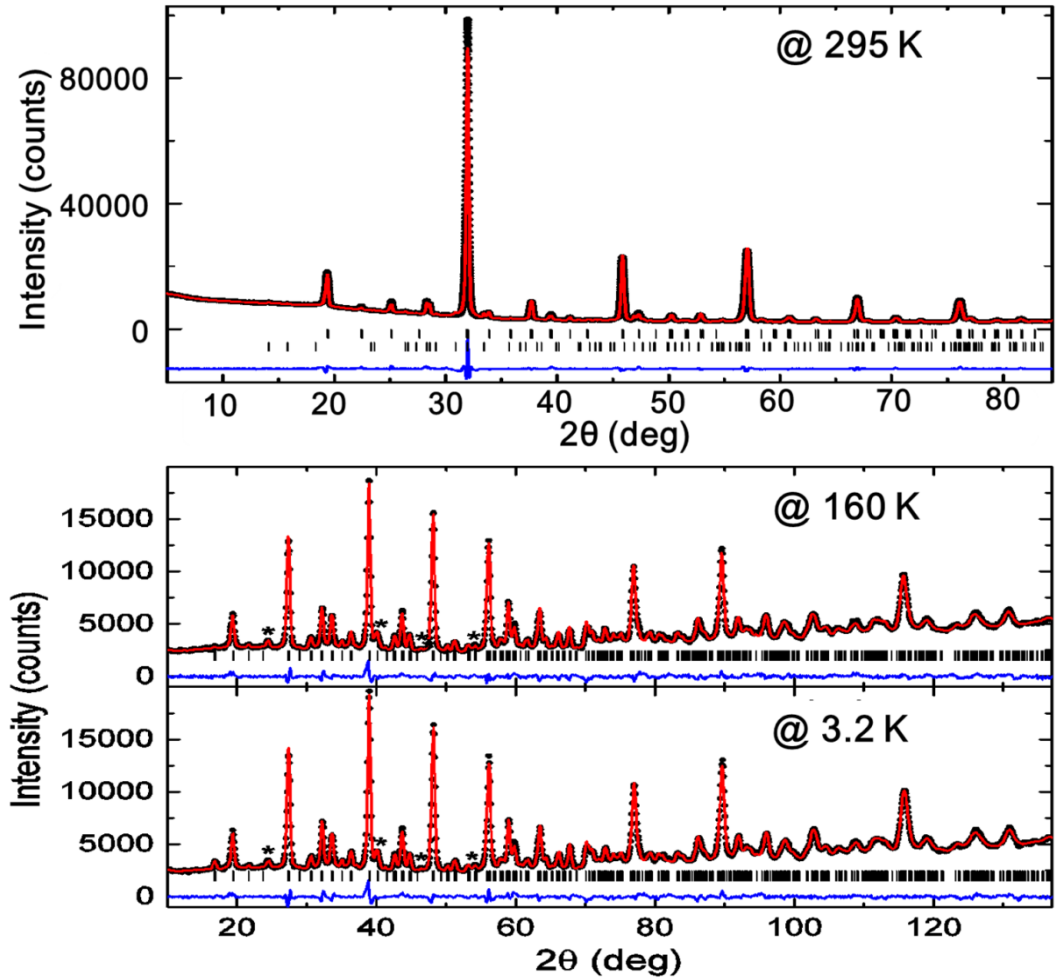

Figure 1. (color online) Top: Rietveld refinement of the *x*-ray powder diffraction data of $La_2Ni_{1.19}Os_{0.81}O_6$ collected at 295 K. The vertical bars indicate the positions of the nuclear Bragg reflections for $La_2Ni_{1.19}Os_{0.81}O_6$ and impurity phase $La_3OsO_7$. Bottom: Rietveld refinements of the powder neutron diffraction data of $La_2Ni_{1.19}Os_{0.81}O_6$ collected at 160 K and 3.2 K (bottom). The strongest peaks of the impurity phase are marked with an asterisk.

Due to the fact that the *x*-ray scattering power of the O atoms is relatively weak in comparison to those of the much heavier La and Os atoms, the crystal structure was also investigated by neutron diffraction at 3.2 and 160 K, well below and above the magnetic ordering temperature of about 125 K. The crystal structure of $La_2Ni_{1.19}Os_{0.81}O_6$ could be successfully refined in the space group $P2_1/n$ with residuals $R_F = 0.0273$ (3.2 K) and $R_F = 0.0250$ (160 K), respectively (Figure 1, bottom). During the refinement, the occupancies of the Ni and Os atoms at the 2c and 2d site were fixed to the values obtained from the x-ray data. Concerning the impurity phase $La_3OsO_7$, only the overall scale factor and the lattice parameters were simultaneously allowed to vary during the Rietveld refinements, whereas the structural parameters obtained at lower temperature (taken from Ref. 26) were fixed. The obtained crystallographic data are summarized in Table I, and the atomic positions are summarized in Table II.

TABLE I. Crystallographic data obtained from refinements of powder x-ray and neutron diffraction data of $La_2Ni_{1.19}Os_{0.81}O_6$ at different temperatures.

|  | 295 K | 160 K | 3.2 K |
|---|---|---|---|
| Diffraction source | x-ray | Neutron | Neutron |
| Space group | $P2_1/n$ | $P2_1/n$ | $P2_1/n$ |
| Lattice parameters | $a = 5.58129(8)$ Å | $a = 5.5742(5)$ Å | $a = 5.5700(4)$ Å |
|  | $b = 5.61271(9)$ Å | $b = 5.6116(5)$ Å | $b = 5.6123(4)$ Å |
|  | $c = 7.89614(12)$ Å | $c = 7.8907(7)$ Å | $c = 7.8866(6)$ Å |
|  | $\beta = 90.075(4)°$ | $\beta = 90.01(3)°$ | $\beta = 90.04(2)°$ |
| Cell volume | 247.356(7) Å$^3$ | 246.82(4) Å$^3$ | 246.54(3) |
| $R_F$ | 0.0180 | 0.0250 | 0.0273 |

TABLE II. The atomic positions and isotropic temperature factors ($B_{iso}$) of $La_2Ni_{1.19}Os_{0.81}O_6$. The refinements were carried out in the monoclinic space group $P2_1/n$. The 2c sites (½,0,½) are occupied by Ni1/Os1 atoms while the 2d sites(0,0,½) are occupied by Ni2 atoms with the occupancy fixed to be 1. The $B_{iso}$ were constrained to be equal for the metal atoms and the oxygen atoms, respectively. All parameters marked with an asterisk were not allowed to vary during the refinements.

| $T$ (K) | 295 | 160 | 3.2 |
|---|---|---|---|
| Diffraction source | x-ray | Neutron | Neutron |
| x (La) | –0.0060(5) | 0.0157(7) | 0.0162(6) |
| y (La) | 0.0379(1) | 0.0441(3) | 0.0441(3) |
| $z$ (La) | 0.2521(3) | 0.2508(19) | 0.2496(18) |
| $x$ (O1) | 0.270(3) | 0.2843(17) | 0.2833(13) |
| $y$ (O1) | 0.295(3) | 0.2809(13) | 0.2850(12) |
| $z$ (O1) | 0.031(5) | 0.0460(11) | 0.0468(10) |
| $x$ (O2) | 0.213(3) | 0.1970(17) | 0.1961(13) |
| $y$ (O2) | 0.796(2) | 0.8001(12) | 0.7971(10) |
| $z$ (O2) | 0.026(5) | 0.0347(11) | 0.0370(10) |
| $x$ (O3) | 0.904(3) | 0.9218(11) | 0.9218(10) |
| $y$ (O3) | 0.481(1) | 0.4882(9) | 0.4846(8) |
| $z$ (O3) | 0.245* | 0.2478(12) | 0.2475(10) |
| $B_{iso}$ (La/Ni/Os) | 0.16(2) | 0.34(2) | 0.25(2) |
| $B_{iso}$ (O) | 1* | 0.55(3) | 0.46(3) |
| $Occ$(Ni1/Os1) | 0.187(3)/0.813(3) | 0.187/0.813* | 0.187/0.813* |

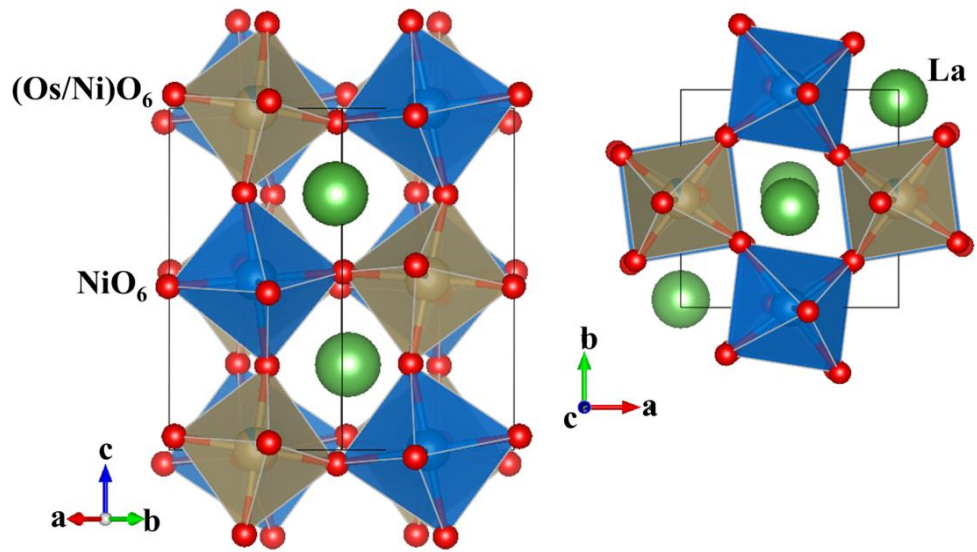

Figure 2. (color online) Crystal structure of $La_2Ni_{1.19}Os_{0.81}O_6$. Blue octahedra are fully centered by $Ni^{2+}$ ions, grey octahedra contain a majority fraction of Os ($Os^{4+}$, $Os^{5+}$) ions and a minor fraction of $Ni^{2+}$ ions.

The inter-octahedral Ni−O−Os bond angles of $La_2Ni_{1.19}Os_{0.81}O_6$ deviate strongly from 180° (Table III), indicating strong octahedral tilting as shown in Fig. 2. The bond distances in the $NiO_6$ and $OsO_6$ octahedra are summarized in Table 3. At 295 K, the average Ni−O bond length, 2.06 Å, is comparable to the literature values for $Ni^{2+}$ double perovskite oxides $A_2NiOsO_6$ ($A$ = Ca, Sr, Ba) [6,27], which indicates that Ni is divalent in this sample.

TABLE III. Selected bond lengths and bond angles of $La_2Ni_{1.19}Os_{0.81}O_6$ derived from *x*-ray and neutron powder patterns at different temperatures.

| Bond length (Å) | 295 K | 160 K | 3.2 K |
|---|---|---|---|
| Ni–O1 | 2.109(16) × 2 | 2.016(8) × 2 | 2.038(7) × 2 |
| Ni–O2 | 1.981(16) × 2 | 2.046(9) × 2 | 2.061(7) × 2 |
| Ni–O3 | 2.086(5) × 2 | 2.038(13) × 2 | 2.040(9) × 2 |
| Os–O1 | 1.912(16) × 2 | 2.038(9) × 2 | 2.021(7) × 2 |
| Os–O2 | 2.052(15) × 2 | 2.029(8) × 2 | 2.014(7) × 2 |
| Os–O3 | 2.011(5) × 2 | 2.004(13) × 2 | 2.002(9) × 2 |
| Ni–O1–Os | 159.6(6) | 154.6(3) | 154.0(3) |
| Ni–O2–Os | 157.8(6) | 152.1(3) | 151.9(3) |
| Ni–O3–Os | 149.1(5) | 154.8(2) | 154.6(2) |

## *B*. X-ray absorption spectroscopy

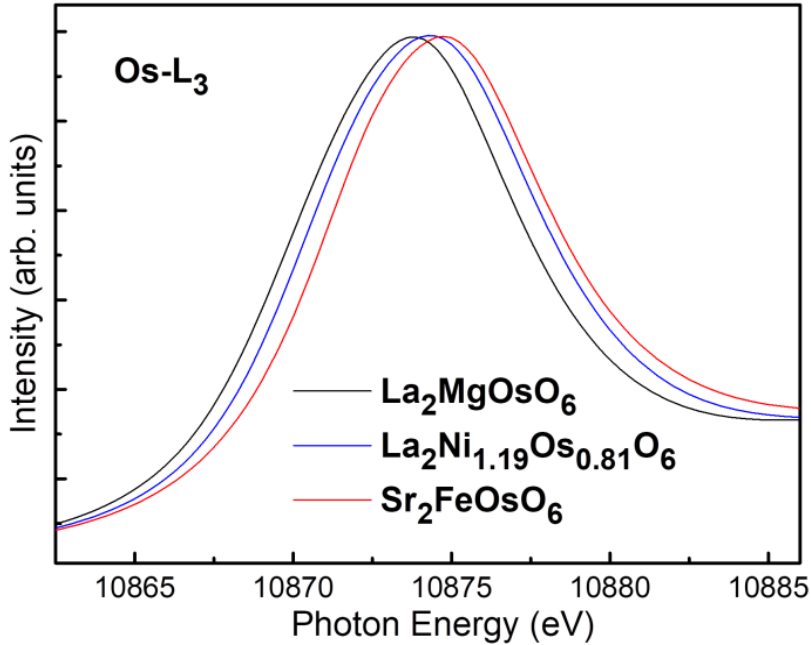

Figure 3. (color online) The Os-$L_3$ XAS spectra of $La_2Ni_{1.19}Os_{0.81}O_6$ and of $Sr_2FeOsO_6$ as $Os^{5+}$ reference and $La_2MgOsO_6$ as $Os^{4+}$ reference.

To explore the valence state of Os, the Os-$L_3$ XAS spectrum of $La_2Ni_{1.19}Os_{0.81}O_6$ was recorded at room temperature. It is well known that XAS spectra are highly sensitive to the valence state: an increase of the valence state of the metal ion by one causes a shift of the XAS $L_{2,3}$ spectra by one or more eV toward higher energies [28,29]. Figure 3 shows the Os-$L_3$ XAS spectrum of $La_2Ni_{1.19}Os_{0.81}O_6$ together with spectra of $Sr_2FeOsO_6$ and $La_2MgOsO_6$ as $Os^{5+}$ and $Os^{4+}$ references, respectively [6]. The energy position of the $La_2Ni_{1.19}Os_{0.81}O_6$ spectrum is located exactly in the middle between that of $Sr_2FeOsO_6$ and $La_2MgOsO_6$ demonstrating the oxidation state of $Os^{4.5+}$. This is consistent with the expected valence state $Os^{4.46+}$ for the composition $La_2Ni_{1.19}Os_{0.81}O_6$ obtained from the Rietveld refinements of the XRD data.

**C. Electrical Transport**

The electrical resistivity of a sintered polycrystalline sample of $La_2Ni_{1.19}Os_{0.81}O_6$ is displayed in Figure 4. The sample shows insulating behavior as the resistivity increases by several orders of magnitude as the temperature decreases and exceeds the measurement limit for temperatures lower than 100 K. The data was plotted on $T^{-1}$ and $T^{-1/4}$ scales, and the plot is found to be roughly linear on a $T^{-1/4}$ scale (see the inset of Fig. 4), in accordance with a three-dimensional variable range hopping transport model.

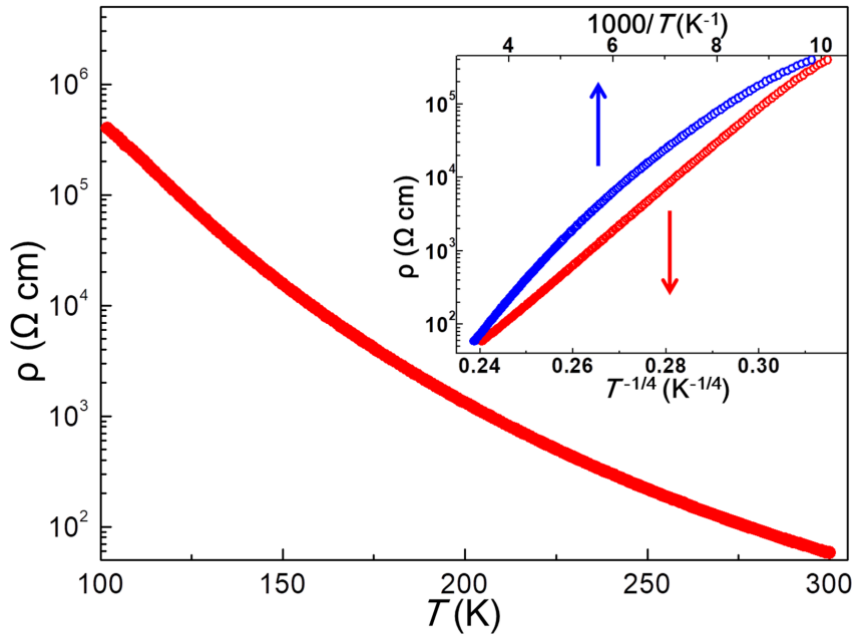

Figure 4. (color online) Temperature dependence of the electrical resistivity of $La_2Ni_{1.19}Os_{0.81}O_6$. The inset shows the corresponding plots on $T^{-1}$ and $T^{-1/4}$ scales.

**D. Magnetic properties**

The temperature dependence of the magnetic susceptibility χ of La$_2$Ni$_{1.19}$Os$_{0.81}$O$_6$ is shown in Figure 5. The sharp increase of χ below about 125 K with cooling indicates a possible ferro- or ferrimagnetic transition. Because the minor impurity La$_3$OsO$_7$ is antiferromagnetic with a Néel temperature of 45 K [26], the sharp transition around 125 K is intrinsic for La$_2$Ni$_{1.19}$Os$_{0.81}$O$_6$. The convex $\chi^{-1}$ vs $T$ curve above the magnetic transition temperature indicates a ferrimagnetic transition. An attempt to fit the data above 400 K with the Curie-Weiss law resulted in a Curie-Weiss temperature ($\theta_{CW}$) of −73 K and an effective magnetic moment ($\mu_{eff}$) per formula unit (f.u.) of 3.63 μ$_B$. The negative $\theta_{CW}$ indicates that antiferromagnetic interactions are dominant in this sample, supporting the ferrimagnetic phase transition. The obtained $\mu_{eff}$ compares reasonably well with $\mu_{eff}$ = 3.69 μ$_B$/f.u., which is calculated from the formula La$_2$Ni$_{1.19}^{2+}$Os$_{0.43}^{4+}$Os$_{0.38}^{5+}$O$_6$ if one assumes a spin-only moment for Ni$^{2+}$ (2.83 μ$_B$), and takes the average $\mu_{eff}$ = 3.28 μ$_B$ for Os$^{5+}$ as obtained for some Os$^{5+}$ double perovskites [30–33]. Any possible Van-Vleck type paramagnetic contribution to $\mu_{eff}$ of the Os$^{4+}$ ions is neglected in this rough estimate.

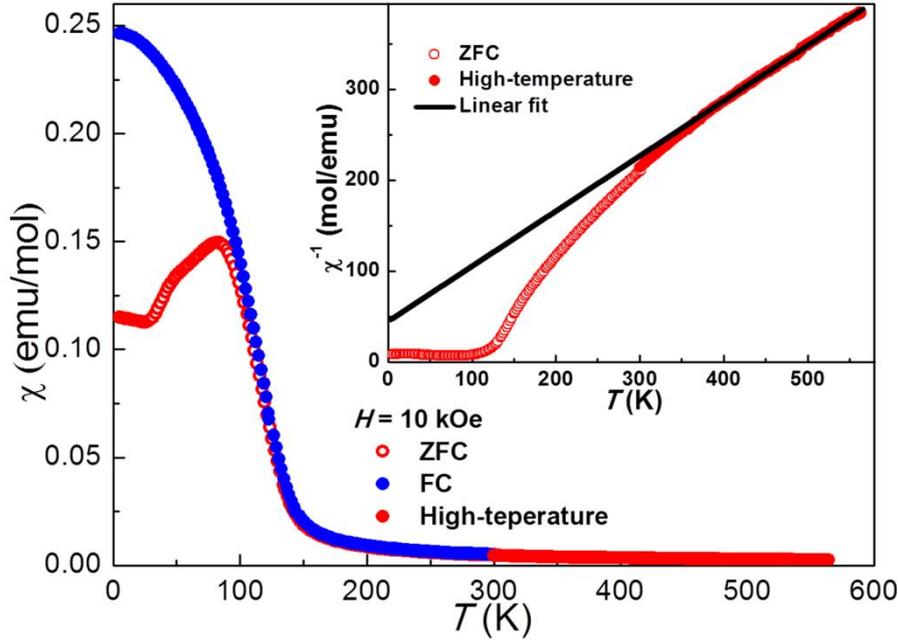

Figure 5. (color online) Temperature dependence of the magnetic susceptibility of La$_2$Ni$_{1.19}$Os$_{0.81}$O$_6$. The inset shows the corresponding $\chi^{-1}$ vs $T$ plot.

To further characterize the magnetic transition, the isothermal magnetization curves of La$_2$Ni$_{1.19}$Os$_{0.81}$O$_6$ were measured above and below the transition temperature after zero field cooling. The linear behavior of the $M(H)$ data at 200 K is consistent with the paramagnetic state. At 5 K, the $M(H)$ curve collected between −50 kOe and +50 kOe shows a large hysteresis, but the hysteresis loop cannot be closed. Thus, the $M(H)$ curve was measured between −140 kOe and +140 kOe, see Figure 6. The magnetization is about 0.65 $\mu_B$/f.u. at 5 K and 140 kOe, but still not saturated. A remarkable feature of the $M(H)$ curve of $La_2Ni_{1.19}Os_{0.81}O_6$ is the high coercive field $H_C$ = 41 kOe at $T$ = 5 K. The $M(H)$ curve was also measured after field cooling (50 kOe) and found to be identical to that measured after zero field cooling, no exchange bias effect is observed.

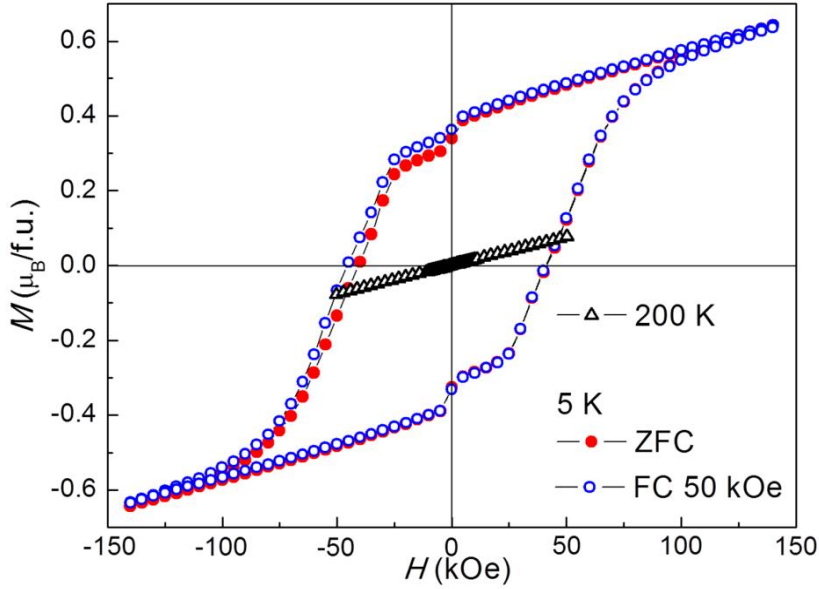

Figure 6. (color online) The isothermal magnetization of $La_2Ni_{1.19}Os_{0.81}O_6$ measured at 200 K and 5 K. The small step near $H$ = 0 may be due to a soft magnetic impurity.

**E. Magnetic structure**

In order to investigate the magnetic structure of $La_2Ni_{1.19}Os_{0.81}O_6$, we have collected several neutron powder patterns on the instrument E6 of BER II ($\lambda$ = 2.42 Å) from 1.7 up to 134 K. In agreement with our magnetization measurements a magnetic contribution to the powder patterns could be observed below 125 K. The strongest magnetic intensity was observed at the position of the reflections 011, 101 and 10−1 (Fig. 7). For the Ni and Os atoms at the positions 2$d$(½,0,0; 0,½,½) and 2$c$(½,0,½; 0,½,0) magnetic intensity could be generated with a model, where the spins located at both sites are coupled antiparallel with a spin sequence − − + +. In the pattern, shown in Fig. 7, additional weak intensities were observed at 2θ = 16.7 and 24.1 °, which are close to the position of the reflection 001, and that of the reflection pair 100/010 (forbidden in $P2_1/n$), respectively. The third reflection observed at 21.5 °can be indexed as (½½1)$_M$ belonging to the impurity $La_3OsO_7$ [26]. In order to characterize these reflections in more detail, we have collected powder patterns at 1.7 and 150 K on the instrument E2 with a much higher counting rate and better instrumental resolution. From the difference pattern shown in the insert of Fig. 7 magnetic intensity was clearly detected for reflection 001, whereas for the reflection pair 100 and 010 the magnetic contribution was found to be negligible. Therefore the intensity observed at 24.1 ° should belong to another unknown impurity.

From our *x*-ray diffraction study, it was found that the osmium site 2*c* site is partially occupied with Ni atoms giving a ratio *occ*(Ni1)/*occ*(Os1) = 0.187(3)/0.813(3), while the 2*d* site is fully occupied by Ni2. For the determination of the moment values from the refinements of the magnetic structure we have assumed that Ni1 and Ni2, located at 2*c* and 2*d* sites, have the same magnetic moment, while initially the Os moments were kept at zero. Further, we used the positional and thermal parameters as well as the monoclinic angle *β* as determined more precisely from the analysis of the E9 data. In order to determine the moment direction, we carried out Rietveld refinements using model structures. In the case of an ordering along the *b* and *c* axes, it could be seen that the magnetic peak significantly shifted to a higher 2θ value, while a satisfactory fit was obtained when the moments are aligned to the *a* axis. In fact, due to the lattice distortions the reflection pair 101/10−1 is shifted to a higher 2θ value in comparison to the 011 (see inset of Fig. 7). From our calculations, we obtained the following intensity ratios *I*(011/01−1)/*I*(101/10−1): ~3/1 (μ // *a*), ~1/3 (μ // *b*), ~1/1 (μ // *c*). This shows that the best calculated peak position is reached when the strongest magnetic intensity appears on the reflection 011 generated with a moment direction parallel to the *a* axis. In the next step, we allowed varying the magnetic moment of the Os atoms at the site 2*c*. Independently from the models used above the refined Os moment was found to be 0.00(6) $\mu_B$. Therefore in the final refinement, the Os moment was fixed to be zero.

In order to generate magnetic intensity at the position of the reflection 001 two models were considered, where the Ni and mixed Os/Ni sites (½,0,0), (0,½,½), (½,0,½) and (0,½,0) have an additional *b* spin component with the equences + − − + and + − + − , respectively. The refinements showed that the spin sequence + − + − results in similarly calculated intensities for the reflections 100 and 001. But the inset of Fig. 7 shows that the magnetic intensity of the 100 is much weaker than that of the 001 reflection. A reasonable fit was obtained by using the spin sequence + − − +. Together with the spin component along the *a* direction with the spin sequence − − + + one obtains a spin structure which is noncollinear in the *ab* plane, but has collinear spin chains along the *c* direction (Fig. 8).

This spin arrangement is compatible with the crystal structure symmetry as shown by a representation analysis [34]. For both atoms at the sites $2d(½,0,0; 0,½,½)$ and $2c(½,0,½; 0,½,0)$ one finds an irreducible representation $\Gamma$ with the spin sequence $+-$ along the $b$ direction and a sequence $++$ along $a$ and $c$. Finally for the 1.7 K data set collected on E2 we obtained the moment values $\mu_x(Ni) = 1.77(5)$ $\mu_B$ and $\mu_y(Ni) = 0.60(12)$ $\mu_B$ resulting in a total moment $\mu_{exp}(Ni) = 1.87(5)$ $\mu_B$ and a residual $R_M = 0.073$ (defined as $R_M = \sum||I_{obs}| - |I_{calc}||/\sum|I_{obs}|$). In this noncollinear magnetic structure, the moments form a tilting angle to the $a$ axis of 19(1) °. For the Rietveld refinements of the powder patterns collected on E6, we have fixed the component $\mu_y(Ni)$. Here we have obtained the component $\mu_x(Ni) = 1.81(5)$ $\mu_B$ and a total moment $\mu_{exp}(Ni) = 1.90(5)$ $\mu_B$, respectively. This value is only slightly smaller than the spin-only moment of 2.0 $\mu_B$ expected for $Ni^{2+}$ ions in octahedral coordination environment ($t_{2g}^6 e_g^2$ electron configuration). The temperature dependence of the obtained Ni moment, shown in Fig. 9, yields an ordering temperature of 125(3) K. Due to the weakness of the magnetic reflection $(001)_M$ we only could follow the temperature dependence of the moment of the much stronger pronounced $x$ component from E6 data.

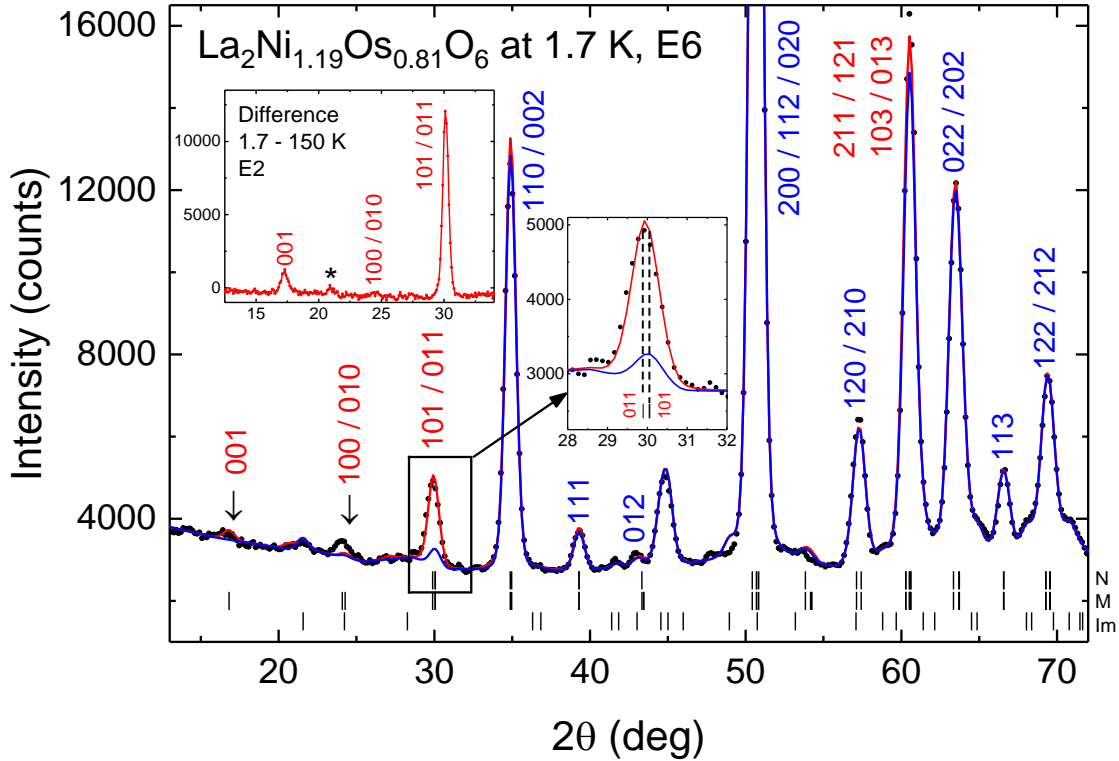

Figure 7. (color online) Neutron powder pattern of $La_2Ni_{1.19}Os_{0.81}O_6$ taken on E6 at 1.7 K. The calculated patterns [nuclear contribution (blue), sum of the nuclear and the magnetic

contribution (red)] are compared with the observations (black-filled circles). The positions (black bars) of the nuclear (N) and magnetic (M) Bragg reflections of $La_2Ni_{1.19}Os_{0.81}O_6$, and those of the impurity phase $La_3OsO_7$ (Im) are shown. The region, where the strongest magnetic reflection $(011)_M$, and the pair $(101)_M$ and $(10-1)_M$ occur is shown in the inset. The second inset (left) shows the presence of the magnetic reflection $(001)_M$ as obtained from the difference pattern (1.7 – 150 K) from E2 data. The weak magnetic intensity observed at 21.5 ° (marked with an asterisk) can be assigned to the magnetic reflection $(½½1)_M$ of the impurity $La_3OsO_7$ [26]. All prominent magnetic reflections are labeled in red color.

Our data analysis suggests that the observed ferrimagnetic structure is driven by the antiferromagnetic interactions between $Ni^{2+}$ ions on the $2d$ sites and the additional $Ni^{2+}$ ions on the $2c$ sites, which leads to a partial cancellation of magnetic moments. It is noted, however, that an unambiguous refinement of Os moments in double perovskites with two magnetic ions at the $B$ and $B'$ site is difficult [5] and thus it cannot be ruled out that ordered $Os^{5+}$ moments contribute to the stabilization of the ferrimagnetic spin structure.

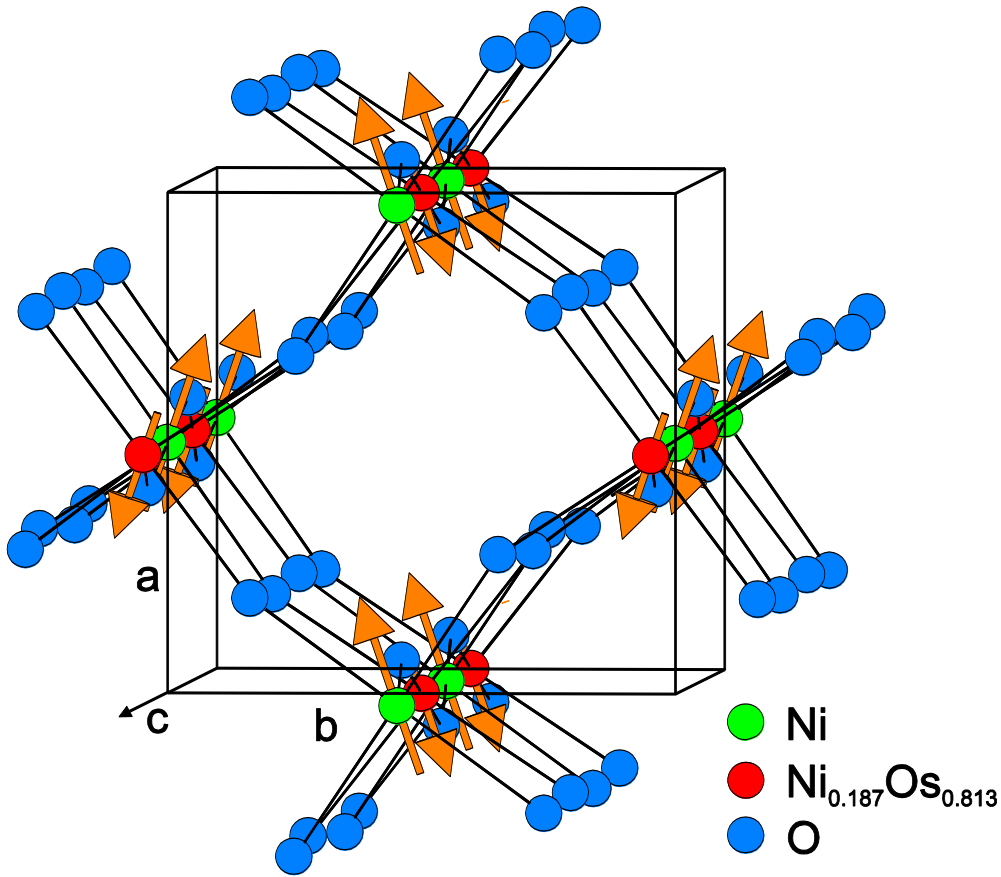

Figure 8. (color online) Magnetic structure of $La_2Ni_{1.19}Os_{0.81}O_6$. For the Ni and mixed Os/Ni sites at $2d(½,0,0; 0,½,½)$ and $2c(½,0,½; 0,½,0)$ only the Ni atoms contribute to the magnetic ordering. The Os site was found to be partially occupied with 19 % Ni. The Ni moments at both sites were constrained to be equal during the Rietveld refinements resulting in a moment $\mu_{exp}(Ni) = 1.90(5)$ $\mu_B$. Therefore the total moment at the Os/Ni site only reached the value $0.19 \times \mu_{exp}(Ni) = 0.36(1)$ $\mu_B$. For clarity, the magnitude of the Ni moment at the Os/Ni site is exaggerated. The magnetic atoms form ferrimagnetic chains along the *c* axis, and within the *ab* plane one finds a noncollinear spin arrangement.

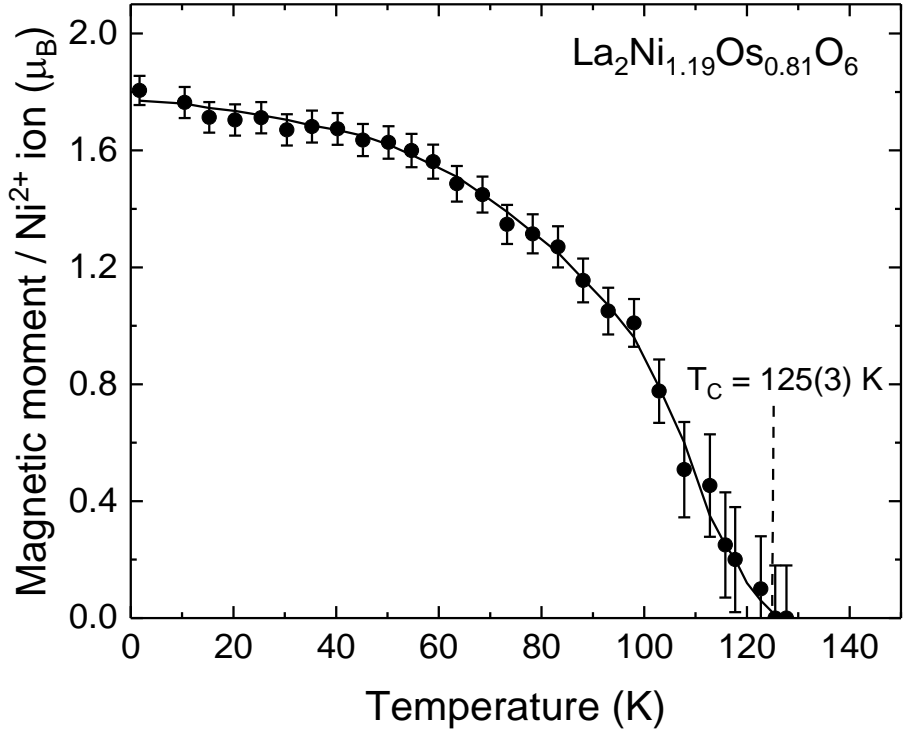

Figure 9. Temperature dependence of the magnetic moments of the $Ni^{2+}$ ions in $La_2Ni_{1.19}Os_{0.81}O_6$. The Wyckoff 2*c* site is occupied with 81 % Os1 and 19 % Ni1, while the 2*d* site only contains Ni2. During the refinements of the magnetic moments, we have applied the constraint $\mu(Ni1) = \mu(Ni2)$, while the magnetic moment of Os was set to be zero. The bold line is a guide for the eye.

## IV. DISCUSSION

We synthesized a non-stoichiometric double perovskite $La_2Ni_{1.19}Os_{0.81}O_6$ containing merely $Ni^{2+}$ ions at the $2d$ sites but a mixed occupancy of paramagnetic ($Ni^{2+}$, $Os^{5+}$) and nonmagnetic ($Os^{4+}$) ions at the $2c$ sites of the monoclinic crystal structure. Formally the compound may be written as $La_2Ni^{2+}(Ni_{0.19}^{2+}Os_{0.43}^{4+}Os_{0.38}^{5+})O_6$. The most remarkable property of this insulating compound is that it features a ferrimagnetic transition with an extraordinary broad hysteresis at 5 K. The spin structure is noncollinear in the *ab* plane.

For rationalizing the magnetic properties we consider first the so far unknown stoichiometric double perovskite $La_2Ni^{2+}Os^{4+}O_6$ with the perfect atomic order, where the magnetism should be entirely determined by antiferromagnetic exchange interactions between the half-filled $e_g$ orbitals of the $Ni^{2+}$ ions as the $Os^{4+}$ ions are expected to be nonmagnetic ($J_{eff} = 0$). Therefore, for $La_2NiOsO_6$ antiferromagnetic ordering is anticipated as is indeed observed for other $A_2NiB'O_6$ compounds with nonmagnetic ions on the $B'$ site ($B'$ = $W^{6+}$, $Ti^{4+}$, $Ir^{5+}$) [35–38], c.f. Table IV. The adopted antiferromagnetic spin structure will be determined mainly by the balance between the nearest neighbor $Ni^{2+}$–O–O–$Ni^{2+}$ interactions and the next nearest neighbor $Ni^{2+}$–O–$Os^{4+}$–O–$Ni^{2+}$ exchange pathways. However, due to the non-stoichiometry the actual compound $La_2Ni_{1.19}Os_{0.81}O_6$ contains a large fraction of paramagnetic $Ni^{2+}$ and $Os^{5+}$ ions at the $2c$ sites, which is the origin for the observed ferrimagnetism with $T_C$ = 125 K. Our analysis of the powder neutron diffraction data indicates that the ferrimagnetism may be solely driven by the antiferromagnetic interactions between the majority of $Ni^{2+}$ ions at the $2d$ sites and the minority of $Ni^{2+}$ ions at the $2c$ sites, which implies that the $Os^{5+}$ moments remain magnetically disordered and possibly freeze at low temperatures. However, this analysis may not be unambiguous and also antiferromagnetic $Ni^{2+}$–O–$Os^{5+}$ interactions may contribute to the stabilization of the noncollinear ferrimagnetic spin structure. The latter interactions are a consequence of the distorted crystal structure. Since the $t_{2g}$ orbitals of $Ni^{2+}$ are completely filled, the only possible antiferromagnetic nearest neighbor exchange pathway is via virtual hopping between half-filled Ni-$e_g$ and Os-$t_{2g}$ orbitals. The Ni-$e_g$ and Os-$t_{2g}$ orbitals are orthogonal in the cubic double perovskite structure, but these interactions become possible in the monoclinic double perovskite where the Ni−O−Os bond angles are strongly reduced from 18o°. This mechanism was invoked for explaining the ferrimagnetic state in $Ca_2Ni^{2+}Os^{6+}O_6$ [42]. Antiferromagnetic interactions between $Ni^{2+}$ ($d^8$) and $Os^{5+}$ ($d^3$) ions appear to be in contradiction to the Goodenough-Kanamori rules which predict a ferromagnetic interaction due to virtual hopping between the half-filled and empty $e_g$ orbitals at the $Ni^{2+}$ and $Os^{5+}$ sites, respectively. However, compared to pure $3d$ systems like

La$_2$NiMnO$_6$ [43,44] the crystal field splitting at the 5$d$ sites is much enhanced and the ferromagnetic coupling involving the σ-exchange pathway between the $e_g$-orbitals becomes weak, in particular in the distorted double perovskite structure [45,46].

Most remarkably the magnetization of the present powder sample of La$_2$Ni$_{1.19}$Os$_{0.81}$O$_6$ features a broad hysteresis with a coercive field as high as 41 kOe. Similar giant coercive fields have been reported for diverse materials classes. Examples include rare earth transition metal films [47], ferrimagnetic transition metal – radical single chain magnets [48], intermetallic systems like Mn$_{1.5}$Ga films [49] as well as ferrimagnetic Fe$_3$Se$_4$ nanostructures [50]. In transition metal oxides extremely high coercivities of 90 and 120 kOe were found for single crystals of the ferrimagnet LuFe$_2$O$_4$ [51] and weakly ferromagnetic Sr$_5$Ru$_{5-x}$O$_{15}$ [52], respectively. Such materials are of interest in the search of rare earth free hard magnets. The detailed mechanism of giant coercivity is not generally understood. However, large magnetocrystalline anisotropy in combination with frustration and/or defects and disorder appears to favor huge coercivities [51]. A large magnetocrystalline anisotropy is indeed expected for La$_2$Ni$_{1.19}$Os$_{0.81}$O$_6$ due to the low-symmetry crystal structure and the presence of the heavy Os atoms. The non-collinear spin arrangement in the $ab$ plane can be attributed to the Dzyaloshinskii-Moriya interaction which induces spin-canting, whereas the high coercivity is possibly associated with a peculiar microstructure and competing exchange interactions introduced by the atomic disorder.

TABLE IV. Comparison of the magnetic properties of La$_2$Ni$_{1.19}$Os$_{0.81}$O$_6$ with those of some Ni$^{2+}$ based double perovskite oxides. The ions on the $B$' site of the double perovskite structure are either non-magnetic (W$^{6+}$, Ti$^{4+}$, Ir$^{5+}$) or magnetic (Os$^{5+}$, Ir$^{6+}$).

| Magnetic ions | composition | Space group | Properties | $T_m$ (K) | $\theta_w$ (K) | $\mu_{eff}$ ($\mu_B$) | Ref. |
|---|---|---|---|---|---|---|---|
| Ni$^{2+}$ | Ca$_2$NiWO$_6$ | $P2_1/n$ | AFM | 52.5 | −75 | 2.85 | 35 |
| Ni$^{2+}$ | Sr$_2$NiWO$_6$ | $I4/m$ | AFM | 54 | −175 | 3.05 | 36 |
| Ni$^{2+}$ | La$_2$NiTiO$_6$ | $P2_1/n$ | AFM | 25 | −60 | 3.12 | 37 |
| Ni$^{2+}$ | SrLaNiIrO$_6$ | $P2_1/n$ | AFM | 74 | −90 | 3.3 | 38 |
| Ni$^{2+}$ + Os$^{4.5+}$ | La$_2$Ni$_{1.19}$Os$_{0.81}$O$_6$ | $P2_1/n$ | FIM | 125 | −73 | 3.63 | This work |
| Ni$^{2+}$ + Os$^{5+}$ | SrLaNiOsO$_6$ | $P2_1/n$ | AFM | 60 | −23 | 4.13 | 39 |
| Ni$^{2+}$ + Os$^{5+}$ | CaLaNiOsO$_6$ | $P2_1/n$ | AFM | 30 | −83 | 3.87 | 40 |
| $Ni^{2+}$ + $Ir^{6+}$ | $Sr_2NiIrO_6$ | $P2_1/n$ | AFM | 58 | - | - | 41 |

## Conclusions

Targeting the $Os^{4+}$ ($J_{eff}$ = 0) compound $La_2NiOsO_6$, we actually obtained nonstoichiometric polycrystalline samples of $La_2Ni_{1.19}Os_{0.81}O_6$ by solid state reaction. As many other double perovskite oxides, $La_2Ni_{1.19}Os_{0.81}O_6$ crystallizes in the monoclinic space group $P2_1/n$. Here, the B sites are fully occupied by $Ni^{2+}$ ions whereas the B' sites feature a mixed occupancy with paramagnetic $Ni^{2+}$ and $Os^{5+}$ as well as non-magnetic $Os^{4+}$ ions. Whereas stoichiometric $La_2NiOsO_6$ is expected to be an antiferromagnet, $La_2Ni_{1.19}Os_{0.81}O_6$ reveals a ferrimagnetic transition at 125 K and adopts a ferrimagnetic spin arrangement with collinear chains along the c axis, but spin canting in the ab plane. Our data analysis indicates that only $Ni^{2+}$ magnetic moments are long-range ordered, but the behavior of the $Os^{5+}$ moments is not entirely clear. The magnetization curve of $La_2Ni_{1.19}Os_{0.81}O_6$ is characterized by a huge coercive field, $H_C$ = 41 kOe at $T$ = 5 K which is attributed to the combined influence of large magnetocrystalline anisotropy and atomic disorder. Our results suggest that careful tuning of the chemical composition may lead to the discovery of new rare earth free hard magnets in the class of double perovskite oxides.

## Acknowledgement

The work in Dresden was partially supported by the Deutsche Forschungsgemeinschaft through SFB 1143. We thank R. Koban, H. Borrmann, and U. Burkhardt for performing magnetization, x-ray, and EDX measurements.